
\font\twelverm=cmr12                    \font\twelvei=cmmi12
\font\twelvesy=cmsy10 scaled 1200       \font\twelveex=cmex10 scaled 1200
\font\twelvebf=cmbx12                   \font\twelvesl=cmsl12
\font\twelvett=cmtt12                   \font\twelveit=cmti12
\font\twelvesc=cmcsc10 scaled 1200      \font\twelvesf=cmss12
                     
\font\twelvemib=cmmib10 scaled 1200
\font\tenmib=cmmib10
\font\eightmib=cmmib10 scaled 800

\skewchar\twelvei='177                  \skewchar\twelvesy='60
\skewchar\twelvemib='177
\newfam\mibfam

\def\twelvepoint{\normalbaselineskip=12.4pt plus 0.1pt minus 0.1pt
  \abovedisplayskip 12.4pt plus 3pt minus 9pt
  \belowdisplayskip 12.4pt plus 3pt minus 9pt
  \abovedisplayshortskip 0pt plus 3pt
  \belowdisplayshortskip 7.2pt plus 3pt minus 4pt
  \smallskipamount=3.6pt plus1.2pt minus1.2pt
  \medskipamount=7.2pt plus2.4pt minus2.4pt
  \bigskipamount=14.4pt plus4.8pt minus4.8pt
  \def\rm{\fam0\twelverm}          \def\it{\fam\itfam\twelveit}%
  \def\sl{\fam\slfam\twelvesl}     \def\bf{\fam\bffam\twelvebf}%
  \def\mit{\fam 1}                 \def\cal{\fam 2}%
  \def\sc{\twelvesc}               \def\tt{\twelvett}%
  \def\sf{\twelvesf}               \def\mib{\fam\mibfam\twelvemib}%
  \textfont0=\twelverm   \scriptfont0=\tenrm   \scriptscriptfont0=\sevenrm
  \textfont1=\twelvei    \scriptfont1=\teni    \scriptscriptfont1=\seveni
  \textfont2=\twelvesy   \scriptfont2=\tensy   \scriptscriptfont2=\sevensy
  \textfont3=\twelveex   \scriptfont3=\twelveex\scriptscriptfont3=\twelveex
  \textfont\itfam=\twelveit
  \textfont\slfam=\twelvesl
  \textfont\bffam=\twelvebf \scriptfont\bffam=\tenbf
			    \scriptscriptfont\bffam=\sevenbf
  \textfont\mibfam=\twelvemib \scriptfont\mibfam=\tenmib
			      \scriptscriptfont\mibfam=\eightmib
  \normalbaselines\rm}

\mathchardef\alpha="710B
\mathchardef\beta="710C
\mathchardef\gamma="710D
\mathchardef\delta="710E
\mathchardef\epsilon="710F
\mathchardef\zeta="7110
\mathchardef\eta="7111
\mathchardef\theta="7112
\mathchardef\iota="7113
\mathchardef\kappa="7114
\mathchardef\lambda="7115
\mathchardef\mu="7116
\mathchardef\nu="7117
\mathchardef\xi="7118
\mathchardef\pi="7119
\mathchardef\rho="711A
\mathchardef\sigma="711B
\mathchardef\tau="711C
\mathchardef\phi="711E
\mathchardef\chi="711F
\mathchardef\psi="7120
\mathchardef\omega="7121
\mathchardef\varepsilon="7122
\mathchardef\vartheta="7123
\mathchardef\varpi="7124
\mathchardef\varrho="7125
\mathchardef\varsigma="7126
\mathchardef\varphi="7127
\def\beginlinemode{\endmode
  \begingroup\parskip=0pt \obeylines\def\\{\par}\def\endmode{\par\endgroup}}
\def\beginparmode{\endmode
  \begingroup \def\endmode{\par\endgroup}}
\let\endmode=\par
{\obeylines\gdef\
{}}
\def\singlespace{\baselineskip=\normalbaselineskip}

\def\oneandahalfspace{\baselineskip=\normalbaselineskip
  \multiply\baselineskip by 3 \divide\baselineskip by 2}
\def\doublespace{\baselineskip=\normalbaselineskip \multiply\baselineskip by 2}

\newcount\firstpageno
\firstpageno=2
\footline={\ifnum\pageno<\firstpageno{\hfil}\else{\hfil\twelverm\folio\hfil}\fi}
\def\toppageno{\global\footline={\hfil}\global\headline
  ={\ifnum\pageno<\firstpageno{\hfil}\else{\hfil\twelverm\folio\hfil}\fi}}
\let\rawfootnote=\footnote              
\def\footnote#1#2{{\rm\singlespace\parindent=0pt\parskip=0pt
  \rawfootnote{#1}{#2\hfill\vrule height 0pt depth 6pt width 0pt}}}
\def\raggedcenter{\leftskip=4em plus 12em \rightskip=\leftskip
  \parindent=0pt \parfillskip=0pt \spaceskip=.3333em \xspaceskip=.5em
  \pretolerance=9999 \tolerance=9999
  \hyphenpenalty=9999 \exhyphenpenalty=9999 }
\def\dateline{\rightline{\ifcase\month\or
  January\or February\or March\or April\or May\or June\or
  July\or August\or September\or October\or November\or December\fi
  \space\number\year}}
\def\received{\vskip 3pt plus 0.2fill
 \centerline{\sl (Received\space\ifcase\month\or
  January\or February\or March\or April\or May\or June\or
  July\or August\or September\or October\or November\or December\fi
  \qquad, \number\year)}}
\hsize=6.5truein
\vsize=8.9truein
\def\nooffset{\global\hoffset=0pt\global\voffset=0pt}

\nooffset
\parskip=\medskipamount
\def\\{\cr}
\twelvepoint            
\doublespace            
\overfullrule=0pt       


\def\preprintno#1{
 \rightline{\rm #1}}    

\def\title                      
  {\null\vskip 3pt plus 0.2fill
   \beginlinemode \doublespace \raggedcenter \bf}

\def\author                     
  {\vskip 3pt plus 0.2fill \beginlinemode
   \singlespace \raggedcenter\rm}

\def\affil                      
  {\beginlinemode\oneandahalfspace \raggedcenter \sl}

\def\abstract                   
  {\vskip 3pt plus 0.3fill \beginparmode
   \oneandahalfspace ABSTRACT: }

\def\endtitlepage               
  {\endpage                     
   \body}
\let\endtopmatter=\endtitlepage

\def\body                       
  {\beginparmode}               

\def\head#1{                    
  \goodbreak\vskip 0.4truein    
  {\immediate\write16{#1}
   \raggedcenter {\sc #1}\par}
   \nobreak\vskip 0truein\nobreak}

\def\itemitemitem{\par\indent\indent \hangindent3\parindent \textindent}
\def\itemitemitemitem{\par\indent\indent\indent \hangindent4\parindent
\textindent}
\def\beginitems{\par\medskip\bgroup
  \def\i##1 {\par\noindent\llap{##1\enspace}\ignorespaces}%
  \def\ii##1 {\item{##1}}%
  \def\iii##1 {\itemitem{##1}}%
  \def\iiii##1 {\itemitemitem{##1}}%
  \def\iiiii##1 {\itemitemitemitem{##1}}
  \leftskip=36pt\parskip=0pt}\def\enditems{\par\egroup}

\def\makefigure#1{\parindent=36pt\item{}Figure #1}

\def\figure#1 (#2) #3\par{\goodbreak\midinsert
\vskip#2
\bgroup\makefigure{#1} #3\par\egroup\endinsert}

\def\beneathrel#1\under#2{\mathrel{\mathop{#2}\limits_{#1}}}

\def\refto#1{$^{#1}$}           

\def\references                 
  {\head{References}            
   \beginparmode
   \frenchspacing \parindent=0pt \leftskip=1truecm
   \parskip=8pt plus 3pt \everypar{\hangindent=\parindent}}

\gdef\refis#1{\item{#1.\ }}                     

\gdef\journal#1, #2, #3, 1#4#5#6{               
    {\sl #1~}{\bf #2}, #3 (1#4#5#6)}            

\def\refstylenp{                
  \gdef\refto##1{ [##1]}                        
  \gdef\refis##1{\item{##1)\ }}                 
  \gdef\journal##1, ##2, ##3, ##4 {             
     {\sl ##1~}{\bf ##2~}(##3) ##4 }}

\def\prd{\journal Phys. Rev. D, }

\def\rmp{\journal Rev. Mod. Phys., }

\def\pl{\journal Phys. Lett., }

\def\endreferences{\body}

\def\figurecaptions             
  {\endpage
   \beginparmode
   \head{Figure Captions}
}

\def\endpage                    
  {\vfill\eject}

\def\endpaper                   
  {\endmode\vfill\supereject}

\def\heading                            
  {\vskip 0.5truein plus 0.1truein      
   \beginparmode \def\\{\par} \parskip=0pt \singlespace \raggedcenter}

\def\subheading                         
  {\vskip 0.25truein plus 0.1truein     
   \beginlinemode \singlespace \parskip=0pt \def\\{\par}\raggedcenter}

\def\tag#1$${\eqno(#1)$$}

\def\align#1$${\eqalign{#1}$$}

\def\aligntag#1$${\gdef\tag##1\\{&(##1)\cr}\eqalignno{#1\\}$$
  \gdef\tag##1$${\eqno(##1)$$}}

\def\endaligntag{}

\def\overset #1\to#2{{\mathop{#2}\limits^{#1}}}
\def\underset#1\to#2{{\let\next=#1\mathpalette\undersetpalette#2}}
\def\undersetpalette#1#2{\vtop{\baselineskip0pt
\ialign{$\mathsurround=0pt #1\hfil##\hfil$\crcr#2\crcr\next\crcr}}}

\def\ref#1{Ref.~#1}                     
\def\Ref#1{Ref.~#1}                     
\def\[#1]{[\cite{#1}]}
\def\cite#1{{#1}}
\def\(#1){(\call{#1})}
\def\call#1{{#1}}
\def\taghead#1{}
\def\frac#1#2{{\textstyle {#1 \over #2}}}
\def\half{{\frac 12}}

\def\12{{1\over2}}

\def\sla{\raise.15ex\hbox{$/$}\kern-.57em}
\def\leaderfill{\leaders\hbox to 1em{\hss.\hss}\hfill}
\def\twiddle{\lower.9ex\rlap{$\kern-.1em\scriptstyle\sim$}}
\def\bigtwiddle{\lower1.ex\rlap{$\sim$}}
\def\gtwid{\mathrel{\raise.3ex\hbox{$>$\kern-.75em\lower1ex\hbox{$\sim$}}}}
\def\ltwid{\mathrel{\raise.3ex\hbox{$<$\kern-.75em\lower1ex\hbox{$\sim$}}}}
\def\square{\kern1pt\vbox{\hrule height 1.2pt\hbox{\vrule width 1.2pt\hskip 3pt
   \vbox{\vskip 6pt}\hskip 3pt\vrule width 0.6pt}\hrule height 0.6pt}\kern1pt}
\def\tdot#1{\mathord{\mathop{#1}\limits^{\kern2pt\ldots}}}
\def\happyface{%
$\bigcirc\rlap{\lower0.3ex\hbox{$\kern-0.85em\scriptscriptstyle\smile$}%
\raise0.4ex\hbox{$\kern-0.6em\scriptstyle\cdot\cdot$}}$}
\def\sadface{%
$\bigcirc\rlap{\lower0.25ex\hbox{$\kern-0.85em\scriptscriptstyle\frown$}%
\raise0.43ex\hbox{$\kern-0.6em\scriptstyle\cdot\cdot$}}$}

\def\pmb#1{\setbox0=\hbox{#1}%
  \kern-.025em\copy0\kern-\wd0
  \kern  .05em\copy0\kern-\wd0
  \kern-.025em\raise.0433em\box0 }

\def\e{{\rm e}}

\def\ln{{\rm ln}}


\def\m@th{\mathsurround=0pt}
\def\leftrightarrowfill{$\m@th \mathord\leftarrow \mkern-6mu
 \cleaders\hbox{$\mkern-2mu \mathord- \mkern-2mu$}\hfill
 \mkern-6mu \mathord\rightarrow$}
\def\overleftrightarrow#1{\vbox{ialign{##\crcr
	\leftrightarrowfill\crcr\noalign{\kern-1pt\nointerlineskip}
	$\hfil\displaystyle{#1}\hfil$\crcr}}}
\catcode`@=11
\newcount\tagnumber\tagnumber=0

\immediate\newwrite\eqnfile
\newif\if@qnfile\@qnfilefalse
\def\write@qn#1{}
\def\writenew@qn#1{}
\def\w@rnwrite#1{\write@qn{#1}\message{#1}}
\def\@rrwrite#1{\write@qn{#1}\errmessage{#1}}

\def\taghead#1{\gdef\t@ghead{#1}\global\tagnumber=0}
\def\t@ghead{}

\expandafter\def\csname @qnnum-3\endcsname
  {{\t@ghead\advance\tagnumber by -3\relax\number\tagnumber}}
\expandafter\def\csname @qnnum-2\endcsname
  {{\t@ghead\advance\tagnumber by -2\relax\number\tagnumber}}
\expandafter\def\csname @qnnum-1\endcsname
  {{\t@ghead\advance\tagnumber by -1\relax\number\tagnumber}}
\expandafter\def\csname @qnnum0\endcsname
  {\t@ghead\number\tagnumber}
\expandafter\def\csname @qnnum+1\endcsname
  {{\t@ghead\advance\tagnumber by 1\relax\number\tagnumber}}
\expandafter\def\csname @qnnum+2\endcsname
  {{\t@ghead\advance\tagnumber by 2\relax\number\tagnumber}}
\expandafter\def\csname @qnnum+3\endcsname
  {{\t@ghead\advance\tagnumber by 3\relax\number\tagnumber}}

\def\equationfile{%
  \@qnfiletrue\immediate\openout\eqnfile=\jobname.eqn%
  \def\write@qn##1{\if@qnfile\immediate\write\eqnfile{##1}\fi}
  \def\writenew@qn##1{\if@qnfile\immediate\write\eqnfile
    {\noexpand\tag{##1} = (\t@ghead\number\tagnumber)}\fi}
}

\def\callall#1{\xdef#1##1{#1{\noexpand\call{##1}}}}
\def\call#1{\each@rg\callr@nge{#1}}

\def\each@rg#1#2{{\let\thecsname=#1\expandafter\first@rg#2,\end,}}
\def\first@rg#1,{\thecsname{#1}\apply@rg}
\def\apply@rg#1,{\ifx\end#1\let\next=\relax%
\else,\thecsname{#1}\let\next=\apply@rg\fi\next}

\def\callr@nge#1{\calldor@nge#1-\end-}
\def\callr@ngeat#1\end-{#1}
\def\calldor@nge#1-#2-{\ifx\end#2\@qneatspace#1 %
  \else\calll@@p{#1}{#2}\callr@ngeat\fi}
\def\calll@@p#1#2{\ifnum#1>#2{\@rrwrite{Equation range #1-#2\space is bad.}
\errhelp{If you call a series of equations by the notation M-N, then M and
N must be integers, and N must be greater than or equal to M.}}\else%
 {\count0=#1\count1=#2\advance\count1
by1\relax\expandafter\@qncall\the\count0,%
  \loop\advance\count0 by1\relax%
    \ifnum\count0<\count1,\expandafter\@qncall\the\count0,%
  \repeat}\fi}

\def\@qneatspace#1#2 {\@qncall#1#2,}
\def\@qncall#1,{\ifunc@lled{#1}{\def\next{#1}\ifx\next\empty\else
  \w@rnwrite{Equation number \noexpand\(>>#1<<) has not been defined yet.}
  >>#1<<\fi}\else\csname @qnnum#1\endcsname\fi}

\let\eqnono=\eqno
\def\eqno(#1){\tag#1}
\def\tag#1$${\eqnono(\displayt@g#1 )$$}

\def\aligntag#1\endaligntag
  $${\gdef\tag##1\\{&(##1 )\cr}\eqalignno{#1\\}$$
  \gdef\tag##1$${\eqnono(\displayt@g##1 )$$}}

\def\eqalignno#1{\displ@y \tabskip\centering
  \halign to\displaywidth{\hfil$\displaystyle{##}$\tabskip\z@skip
    &$\displaystyle{{}##}$\hfil\tabskip\centering
    &\llap{$\displayt@gpar##$}\tabskip\z@skip\crcr
    #1\crcr}}

\def\displayt@gpar(#1){(\displayt@g#1 )}

\def\displayt@g#1 {\rm\ifunc@lled{#1}\global\advance\tagnumber by1
	{\def\next{#1}\ifx\next\empty\else\expandafter
	\xdef\csname @qnnum#1\endcsname{\t@ghead\number\tagnumber}\fi}%
  \writenew@qn{#1}\t@ghead\number\tagnumber\else
	{\edef\next{\t@ghead\number\tagnumber}%
	\expandafter\ifx\csname @qnnum#1\endcsname\next\else
	\w@rnwrite{Equation \noexpand\tag{#1} is a duplicate number.}\fi}%
  \csname @qnnum#1\endcsname\fi}

\def\ifunc@lled#1{\expandafter\ifx\csname @qnnum#1\endcsname\relax}

\let\@qnend=\end\gdef\end{\if@qnfile
\immediate\write16{Equation numbers written on []\jobname.EQN.}\fi\@qnend}

\catcode`@=12
\refstylenp
\catcode`@=11
\newcount\r@fcount \r@fcount=0
\def\refreset{\global\r@fcount=0}
\newcount\r@fcurr
\immediate\newwrite\reffile
\newif\ifr@ffile\r@ffilefalse
\def\w@rnwrite#1{\ifr@ffile\immediate\write\reffile{#1}\fi\message{#1}}

\def\writer@f#1>>{}
\def\referencefile{
  \r@ffiletrue\immediate\openout\reffile=\jobname.ref%
  \def\writer@f##1>>{\ifr@ffile\immediate\write\reffile%
    {\noexpand\refis{##1} = \csname r@fnum##1\endcsname = %
     \expandafter\expandafter\expandafter\strip@t\expandafter%
     \meaning\csname r@ftext\csname r@fnum##1\endcsname\endcsname}\fi}%
  \def\strip@t##1>>{}}

\def\citeall#1{\xdef#1##1{#1{\noexpand\cite{##1}}}}
\def\cite#1{\each@rg\citer@nge{#1}}     

\def\each@rg#1#2{{\let\thecsname=#1\expandafter\first@rg#2,\end,}}
\def\first@rg#1,{\thecsname{#1}\apply@rg}       
\def\apply@rg#1,{\ifx\end#1\let\next=\relax
\else,\thecsname{#1}\let\next=\apply@rg\fi\next}

\def\citer@nge#1{\citedor@nge#1-\end-}  
\def\citer@ngeat#1\end-{#1}
\def\citedor@nge#1-#2-{\ifx\end#2\r@featspace#1 
  \else\citel@@p{#1}{#2}\citer@ngeat\fi}        
\def\citel@@p#1#2{\ifnum#1>#2{\errmessage{Reference range #1-#2\space is bad.}%
    \errhelp{If you cite a series of references by the notation M-N, then M and
    N must be integers, and N must be greater than or equal to M.}}\else%
 {\count0=#1\count1=#2\advance\count1
by1\relax\expandafter\r@fcite\the\count0,%
  \loop\advance\count0 by1\relax
    \ifnum\count0<\count1,\expandafter\r@fcite\the\count0,%
  \repeat}\fi}

\def\r@featspace#1#2 {\r@fcite#1#2,}    
\def\r@fcite#1,{\ifuncit@d{#1}
    \newr@f{#1}%
    \expandafter\gdef\csname r@ftext\number\r@fcount\endcsname%
		     {\message{Reference #1 to be supplied.}%
		      \writer@f#1>>#1 to be supplied.\par}%
 \fi%
 \csname r@fnum#1\endcsname}
\def\ifuncit@d#1{\expandafter\ifx\csname r@fnum#1\endcsname\relax}%
\def\newr@f#1{\global\advance\r@fcount by1%
    \expandafter\xdef\csname r@fnum#1\endcsname{\number\r@fcount}}

\let\r@fis=\refis                       
\def\refis#1#2#3\par{\ifuncit@d{#1}
   \newr@f{#1}%
   \w@rnwrite{Reference #1=\number\r@fcount\space is not cited up to now.}\fi%
  \expandafter\gdef\csname r@ftext\csname r@fnum#1\endcsname\endcsname%
  {\writer@f#1>>#2#3\par}}

\def\ignoreuncited{
   \def\refis##1##2##3\par{\ifuncit@d{##1}%
     \else\expandafter\gdef\csname r@ftext\csname
r@fnum##1\endcsname\endcsname%
     {\writer@f##1>>##2##3\par}\fi}}

\def\r@ferr{\endreferences\errmessage{I was expecting to see
\noexpand\endreferences before now;  I have inserted it here.}}
\let\r@ferences=\references
\def\references{\r@ferences\def\endmode{\r@ferr\par\endgroup}}

\let\endr@ferences=\endreferences
\def\endreferences{\r@fcurr=0
  {\loop\ifnum\r@fcurr<\r@fcount
    \advance\r@fcurr by 1\relax\expandafter\r@fis\expandafter{\number\r@fcurr}%
    \csname r@ftext\number\r@fcurr\endcsname%
  \repeat}\gdef\r@ferr{}\global\r@fcount=0\endr@ferences}


\let\r@fend=\endpaper\gdef\endpaper{\ifr@ffile
\immediate\write16{Cross References written on []\jobname.REF.}\fi\r@fend}

\catcode`@=12

\citeall\refto          
\citeall\ref            %
\citeall\Ref            %
\preprintno{UAHEP9403}
\title{Greens Function for Anti de Sitter Space Gravity}
\vskip 0.2in
\author{Gary Kleppe}
\vskip 0.2in
\affil Department of Physics and %
Astronomy, University of Alabama, Tuscaloosa AL 35487
\abstract{We solve for the retarded Greens function for linearized gravity in
a background with a negative cosmological constant, anti de Sitter space. In
this background, it is possible for a signal to reach spatial infinity in a
finite time. Therefore the form of the Greens function depends
on a choice of boundary condition at spatial infinity. We take as our
condition that a signal which reaches infinity should be lost, not
reflected back. We calculate the Greens function associated with this
condition, and show that it reproduces the correct classical solution
for a point mass at the origin, the anti de Sitter-Schwarzchild solution.}
\endtopmatter

\head{Introduction}

The physics of gravitation at energies small compared to the Planck scale is
well described by Einstein's theory of general relativity. Unfortunately, the
nonrenormalizibility of this theory means that
we cannot use perturbation theory to extract predictions at higher energies.
This means that in this
case, either perturbation theory or general relativity is inapplicable.
It is often believed that to learn anything interesting about
quantum gravity, we must first find a formulation of it which allows us to
make predictions at all orders. Unfortunately, it seems that we are a long
way away from accomplishing this goal.

An alternative approach is to consider quantum gravity
at low energies, and see what might be learned. At low energies there
is no doubt that physics is correctly described by the action functional of
Einstein's gravity, plus an infinite series of counterterms which diverge
in the ultraviolet, but go to zero in the infrared limit\refto{strong}.
This approach tells us, for example,
that the gravitational anomaly of the standard model must cancel, which
gives an observationally verified restriction on the particle content of
the model\refto{ramond}.

This low-energy gravity approach has recently been used by Tsamis and
Woodard to suggest a mechanism for suppression of the
cosmological constant\refto{whosonfirst,weinberg}. They consider quantum
gravity in a de Sitter space background,
the maximally symmetric background for gravity with a positive cosmological
constant. They develop a formalism for doing low-energy perturbation theory
for gravity in this background\refto{structure,mode,getphysical}, and argue
that there is a natural relaxation of the effective cosmological constant
over time\refto{strong}.

In this paper we take the first step toward making a similar analysis for
the case of a negative cosmological constant.
The maximally symmetric background for gravity with a negative cosmological
constant is {\it anti de Sitter space.} In this paper, we determine the
appropriate retarded Greens function for the linearized graviton kinetic
operator on a background which is anti de Sitter everywhere and at all times.
The retarded Greens function $G(x,x')$
allows us to determine the linearized response of the metric to a
distribution of energy-momentum $T(x)$, as
$${\rm Response}=\hbox{ Initial Value Term}+\int d^4x'\ G(x,x')\ T(x')
\eqno(bvp)$$
Furthermore, the average of the retarded and advanced Greens functions gives
the imaginary part of the free propagator for the theory:
$${\rm Im}\left\{i\left[_{\rho\sigma}\Delta^{\alpha\beta}\right](x,x')\right\}
=\half\left\{\left[_{\rho\sigma}G^{\alpha\beta}_{{\rm ret}}\right](x,x')
+\left[_{\rho\sigma}G^{\alpha\beta}_{{\rm adv}}\right](x,x')\right\}\eqno()$$
{}From this, the free Feynman propagator can be deduced, up to terms which are
dependent on the vacuum. This of course is the necessary ingredient in
formulating a theory of Feynman diagrams for anti de Sitter space gravity.

In the first section of this paper (after this introduction), we discuss the
geometry of the anti de Sitter space background, and define coordinate
systems which we will need to use. We do this for an arbitrary number $D$ of
spacetime dimensions. We show that this background has the
peculiar feature that it is possible for an observer to travel from finite to
in
coordinate values (or vice-versa)
in a finite time. Therefore, in solving for \(bvp) one needs to specify what
happens to a signal which reaches infinity\refto{rentacar}. In the next
section we write and solve the gauge fixed equations of motion for gravity
in the anti de Sitter background. In the section after that one, we solve
explicitly for the retarded Greens function of anti de Sitter gravity in four
spacetime dimensions, using
the boundary condition that any signal which reaches infinity should be lost.
In the penultimate section we show that our Greens function gives the right
classical limit for a certain source; specifically that it gives
the anti de Sitter-Schwarzchild solution, up to a coordinate transformation,
when the source $T(x)$ is a single
point mass. Finally we will conclude by discussing where to proceed from here,
in particular how anti de Sitter
space might fit into realistic models of the history of the universe, and
how the analysis in this paper might be modified to accomodate this.

\endpage
\head{Anti de Sitter Space}

Anti de Sitter space (AdS)
in $D$ dimensions can be described in terms of $D+1$
coordinates\footnote*{Notation: Latin indices $i,j,$ etc. are spatial
indices which go from 1 to $D-1$. Greek indicies will run from 0 to $D-1$.}
$X^0, X^i, X^D$, in terms of which the metric is
$$ds^2=-(dX^0)^2+dX^idX^i-(dX^D)^2\eqno(MET)$$
We obtain Anti de Sitter space by restricting these coordinates to obey
$$-(X^0)^2+X^iX^i-X^DX^D=-1/h^2\eqno(hyper)$$
for some constant $h$. In this paper we will set $h=1$ for convenience.

The space AdS contains closed timelike curves
(e.g. the circle $(X^0)^2+ (X^D)^2=1$), along which an observer could travel
and return to his own past. This undesirable feature can be removed by
extending
de Sitter space to its universal covering space (CAdS). This just means
that we introduce an integer $N$ (winding number) which we increment by 1
every time we go around the loop, and we consider different values of $N$
for the same $X$ to be different spacetime locations.

It is of course possible to solve \(hyper) in terms
of a set of $D$ coordinates $x^\mu$. These coordinates will then have a
non-flat metric, given by
$$g_{\mu\nu}=-
{\partial X^0\over\partial x^\mu}{\partial X^0\over\partial x^\nu}
+{\partial X^i\over\partial x^\mu}{\partial X^i\over\partial x^\nu}
-{\partial X^D\over\partial x^\mu}{\partial X^D\over\partial x^\nu}
\eqno(meteqn)$$
One such solution (given here for $D=4$, but easily generalized to any $D$) is
$$X^0=\sqrt{r^2+1}\sin\tau\eqno(Xsol1 a)$$
$$X^1=r\sin\theta\cos\phi\eqno(Xsol1 b)$$
$$X^2=r\sin\theta\sin\phi\eqno(Xsol1 c)$$
$$X^3=r\cos\theta\eqno(Xsol1 d)$$
$$X^4=\sqrt{r^2+1}\cos\tau\eqno(Xsol1 e)$$
The metric in these coordinates can be calculated using \(meteqn). It is
the static and isotropic metric
$$g_{\tau\tau}=-1-r^2\eqno(smet a)$$
$$g_{rr}={1\over 1+r^2}\eqno(smet b)$$
$$g_{\theta\theta}=r^2\eqno(smet c)$$
$$g_{\phi\phi}=r^2\sin^2\theta\eqno(smet d)$$
To obtain de Sitter space, we should take the range of $\tau$ to be 0 to
$2\pi$, and identify all observables at these two boundaries. The covering
space CaDS can be obtained simply by letting $\tau$ take any
real value; it is not necessary to introduce winding numbers when these
coordinates are used.

{}From the metric \(smet) we can show that the Riemann tensor for anti de
Sitter space
$${R^\alpha}_{\beta\gamma\delta}\equiv\Gamma^\alpha_{\beta\delta,\gamma}
-\Gamma^\alpha_{\beta\gamma,\delta}+\Gamma^\alpha_{\gamma\rho}
\Gamma^\rho_{\beta\delta}-\Gamma^\alpha_{\delta\rho}\Gamma^\rho_{\beta\gamma}
\eqno()$$
is
$${R^\alpha}_{\beta\gamma\delta}=-\delta^\alpha_\gamma g_{\beta\delta}
+\delta^\alpha_\delta g_{\beta\gamma}\eqno()$$
{}From this, we can see that the anti de Sitter metric obeys the
gravitational equations of motion
$$R_{\alpha\beta}=\Lambda g_{\alpha\beta}\eqno(aaaahricky)$$
for $\Lambda=-(D-1)$.

The equation of motion of a massless signal is obtained by setting
$ds^2=g_{\mu\nu}dx^\mu dx^\nu$ to zero. For a radially moving signal,
this gives $r=\tan (\tau-\tau_0)$. This shows that a
massless signal
can get from zero to infinity or vice-versa in a finite time, namely
$\frac\pi2$. This means that in CAdS a local event will causally influence
all spatial locations after a certain finite time. The consequence of this
is that the CAdS Greens function depends on the choice of an extra boundary
condition, at $r=\infty$\refto{rentacar}. We will argue that the condition
which works is that any signal which reaches $r=\infty$ should be lost.

The static coordinate system is not the most convenient for the calculations
which are to follow.
{}From \(aaaahricky) we can show that the conformal or Weyl tensor
$$C_{\alpha\beta\gamma\delta}\equiv R_{\alpha\beta\gamma\delta}-\frac1{D-2}
\Bigl(g_{\alpha\gamma}R_{\beta\delta}+g_{\beta\delta}R_{\alpha\gamma}
-g_{\alpha\delta}R_{\beta\gamma}-g_{\beta\gamma}R_{\alpha\delta}\Bigr)$$
$$+\frac1{(D-1)(D-2)}\Bigl(g_{\alpha\gamma}g_{\beta\delta}-g_{\alpha\delta}
g_{\beta\gamma}\Bigr)R\eqno(vile)$$
is identically zero for anti de Sitter space. This fact implies that
there exists conformal coordinate systems in which the anti de Sitter
metric is proportional to the Minkowski metric. One such coordinate
system\def\ibar{{\overline i}} is\footnote*{Barred Latin indices such as
$\ibar$ take values from 1 to $D-2$. This convention and others will be fully
introduced in the next section.}
$t,y,x^\ibar$, in terms of which
$$X^0={t\over y}\eqno(Xsol2 a)$$
$$X^\ibar={x^\ibar\over y},\ \ibar=1,\ldots,D-2\eqno(Xsol2 b)$$
$$X^{D-1}={1+t^2-x^\ibar x^\ibar-y^2\over 2y}\eqno(Xsol2 c)$$
$$X^D={1-t^2+x^\ibar x^\ibar+y^2\over 2y}\eqno(Xsol2 d)$$
We then find
$$g_{\mu\nu}=\frac1{y^2}\eta_{\mu\nu}\eqno()$$

These conformal coordinates do not naturally extend to CAdS; winding numbers
$N$ need to be introduced. The two regions $y>0$ and $y<0$ actually
correspond to different coordinate patches.
It is not possible to go from positive $y$
to negative $y$ by passing through $y=0$, but it is possible to do so by
going through $y=\infty$.
$y=0$ is not really part of the space since no finite values of
the $X$ coordinates make $y$ zero.
To emphasize the separation between
positive and negative $y$ regions, we will take the
convention that regions of $y>0$ are given integer winding numbers, while
regions with $y<0$ are given half integer numbers.

Now consider a massless signal originating at $x'{}^\mu$ on the patch with
winding number $N$. By setting $ds^2=0$, we see that
the signal will follow the usual path of a
lightlike beam, the path $x$ such that $(x-x')^2=0$. If the signal is
initially headed toward $y=0$, then it will at some time hit $y=0$. If it is
initially headed away from $y=0$, then it will go into the far future of
sheet $N$, and reappear in the far past on the sheet $N+\half$. On this
sheet it will be heading toward $y=0$. Our boundary conditions will be that
any signal which hits $y=0$ is lost. Therefore a massless signal from sheet
$N$ can at most get to sheet $N+\half$ before it is lost. Massive (timelike)
signals, on the other hand, can remain present for all $N$. These facts will
be important in the construction of the Greens function.

We see that an observer at $x^\mu$ on sheet $N$ can causally observe any
event $x'{}^\mu$ on sheet $N'=N$ for
$t'<t-\sqrt{(x-x')^\ibar (x-x')^\ibar+(y-y')^2}$;
any event on sheet $N'=N-\half$ for
$t'<t+\sqrt{(x-x')^\ibar (x-x')^\ibar+(y-y')^2}$;
and any event on sheets $N'<N-\half$ no matter where or when. This is the
same effect we saw using static coordinates. Although in conformal coordinates
it takes an infinite amount of time $t$ to cross to another $N$ level, it is
still possible for a massless signal to cross the entire spatial extent of
the manifold without taking the entire temporal lifetime of the manifold to
do it.

The condition \(hyper), and hence the anti de Sitter metric,
is invariant under the {\it anti de Sitter group,} the $D(D+1)/2$ parameter
group of Lorentz rotations of the $X$ coordinates.
In static coordinates, $\tau$ translation and spatial rotations which shift
$\theta,\ \phi$ are some of the anti de Sitter transformations; the others
are nonlinearly realized in these coordinates. In conformal coordinates,
the anti de Sitter symmetries are realized as translational
invariance in the $D-1$ dimensional flat subspace, Lorentz invariance in this
subspace, invariance under simultaneous dilatation of all coordinates, and
$D-1$ nonlinear symmetries. This is exactly analogous to the de Sitter
case\refto{me}.

An important invariant under
the anti de Sitter group is the distance function
$$1-z(X,X')\equiv-\frac14 (X-X')^2\eqno(dist)$$
In the static coordinates this is
$$1-z(x,x')=\half+\half\sqrt{r^2+1}\sqrt{r'{}^2+1}\cos (\tau-\tau')
+\frac{rr'}2
(\sin\theta\sin\theta'\cos(\phi-\phi')+\cos\theta\cos\theta')\eqno(dist2)$$
In conformal coordinates it is
$$1-z(x,x')=-\frac1{2yy'}(x-x')^2\eqno(confdist)$$
Setting $1-z=0$ in either coordinate system is an alternate way to find the
path of a massless signal. Since our Greens function should be invariant
under the anti de Sitter group (at least up to gauge transformations), it will
naturally be a function of $1-z$.

\head{Gauge Fixing and Classical Solutions}
\def\lag{{\cal L}}
We now expand the gravitational Lagrangian
$$\lag={1\over\kappa^2}\sqrt{-g}\left(R-(D-2)\Lambda\right)\eqno()$$
about the anti de Sitter background
$$g_{\mu\nu}=\frac1{y^2}\left(\eta_{\mu\nu}+\kappa\psi_{\mu\nu}\right)\eqno()$$
The resulting Lagrangian for $\psi$
is invariant under the gauge transformation
$$\delta\psi_{\mu\nu}=-2e_{(\mu,\nu)}+\frac2y\eta_{\mu\nu}e_y\eqno(gt)$$
provided that the functions $e_\mu(x)$ are well-behaved enough at infinity
to allow integration by parts.

We borrow the following notation from \refto{structure} (some of which is
standard): Indices are raised and lowered by the (spacelike) Minkowski metric
$\eta_{\mu\nu}$ (since we have given up general covariance by working
exclusively in the conformal coordinate system).
Greek indices go from 0 to $D-1$, while Latin indices take only
spacelike values 1 to $D-1$. Indices in parentheses are
symmetrized ($a_{(\mu\nu)}\equiv \half (a_{\mu\nu}+a_{\nu\mu})$).
A bar over a $\delta$ or $\eta$ tensor indicates the suppression of
its non-flat (in this case $y$) components; e.g. $\overline\eta_{\mu\nu}=
\eta_{\mu\nu}-\delta_\mu^y\delta_\nu^y$.
We also introduce the new notation that barred indices only run over flat
coordinates. Thus a tensor with a barred index \def\mubar{{\overline\mu}}
such as $A^\mubar$ must have zero $y$ component.

Following \refto{structure}, we
expand out the quadratic part of the Lagrangian, and add the gauge fixing
term $-\half F_\mu F_\nu \eta_{\mu\nu}$
where
$$F_\mu\equiv y^{1-D/2}\left(\psi^\nu_{\mu,\nu}-\half\psi_{,\mu}-\frac{D-2}y
\psi_{\mu y}\right)\eqno()$$
Then the gauge fixed quadratic Lagrangian is
$$\lag^{(2)}_{GF}=\half \psi^{\mu\nu} {D_{\mu\nu}}^{\rho\sigma}\psi_{\rho
\sigma}\eqno()$$
where
$${D_{\mu\nu}}^{\rho\sigma}\equiv
\left(\half\delta_\mu^{(\rho}\delta_\nu^{\sigma)}
-\eta_{\mu\nu}\eta^{\rho\sigma}\right)D_0+{D-2\over y^D}
\delta^y_{(\mu}\delta_{\nu)}^{(\rho}\delta^{\sigma)}_y\eqno()$$
with
$$D_0\equiv y^{2-D}\left(\partial^2-{D-2\over y}{\partial\over\partial y}
\right)\eqno()$$
We will now investigate solutions of the homogeneous equations of motion.
The gauge-invariant equations are
$${D_{\mu\nu}}^{\rho\sigma}\psi_{\rho\sigma}+\frac1yF_{(\mu,\nu)}-
\frac1{2y}\eta_{\mu\nu}{F^\rho}_{,\rho}+\frac{D-2}{2y^2}\delta_{(\mu}^y
F_{\nu)}+\frac{D-2}{4y^2}\eta_{\mu\nu}F^y=0\eqno()$$
If we demand $F=0$, then solutions of the gauge-fixed equations
$(D\psi)_{\mu\nu}=0$ are also solutions of the invariant equations. Let us
look for such solutions.
Following the analogous treatment of the $\Lambda>0$ case\refto{mode}, it is
convenient to reexpress $\psi$:
\def\nubar{{\overline\nu}}
$$\psi_{\rho\sigma}=\left(\overline\delta_{(\rho}^\mubar
\overline\delta_{\sigma)}^\nubar-\frac1{D-3}\overline\eta_{\rho\sigma}
\overline\eta^{\mubar\nubar}\right)
\psi^A_{\mubar\nubar}+2\delta^y_{(\rho}\overline\delta_{\sigma)}^\mubar
\psi^B_\mubar
+\left(\frac1{D-3}\overline\eta_{\rho\sigma}-\delta^y_{(\rho}
\delta^y_{\sigma)}\right)\psi^C\eqno(expansion)$$
where $\psi^A$ is symmetric. We then calculate
\def\albar{{\overline\alpha}}\def\bbar{{\overline\beta}}
$${D_{\mu\nu}}^{\rho\sigma}\psi_{\rho\sigma}=\half\left(
\overline\delta^\albar_{(\mu}\ \overline\delta^\bbar_{\nu)}
+\frac1{D-3}\ \delta_\mu^y\ \delta_\nu^y\ \overline\eta^{\albar\bbar}\right)
D_0\ \psi^A_{\albar\bbar}\eqno()$$
$$\qquad+\delta^y_{(\mu}\overline\delta^\albar_{\nu)}\ D_{D-2}\ \psi^B_\albar
-\half
\left(\frac{D-2}{D-3}\right)\delta_\mu^y\delta_\nu^y\ D_{2(D-3)}\ \psi_C$$
where
$$D_n\equiv D_0+{n\over y^D}\eqno(ddef)$$
Thus the equations of motion for the $A,B,C$ components are
$$D_0\ \psi^A_{\mubar\nubar}=0\eqno(dp a)$$
$$D_{D-2}\ \psi^B_\mubar=0\eqno(dp b)$$
$$D_{2(D-3)}\ \psi^C=0\eqno(dp c)$$
The general solutions to these equations are
$$\psi^A_{\mubar\nubar}
=\int{d^{D-2}k\over(2\pi)^{D-1}}\int_{-\infty}^\infty{d\alpha\over2\pi}
\ \e^{ik^\ibar x^\ibar-i\sqrt{\alpha^2+k^2}t}\ \alpha y^{D/2}
h_{{D-2\over2}}(\alpha y)A_{\mubar\nubar}(\alpha,k^\ibar)\hbox{ + c.c.}
\eqno(gensol a)$$
$$\psi^B_\mubar
=\int{d^{D-2}k\over(2\pi)^{D-1}}\int_{-\infty}^\infty{d\alpha\over2\pi}
\ \e^{ik^\ibar x^\ibar-i\sqrt{\alpha^2+k^2}t}\ \alpha y^{D/2}
h_{{D-4\over2}}(\alpha y)B_\mubar(\alpha,k^\ibar)\hbox{ + c.c.}
\eqno(gensol b)$$
$$\psi^C
=\int{d^{D-2}k\over(2\pi)^{D-1}}\int_{-\infty}^\infty{d\alpha\over2\pi}
\ \e^{ik^\ibar x^\ibar-i\sqrt{\alpha^2+k^2}t}\alpha y^{D/2}
h_{{D-6\over2}}(\alpha y)C(\alpha,k^\ibar)\hbox{ + c.c.}
\eqno(gensol c)$$
where $h_l$ are spherical Hankel functions\footnote*{Specifically, they are
spherical Hankel functions of the first kind. Their complex conjugates
$h_l^*$ are Hankel functions of the second kind.}
$$h_l(x)\equiv(-x)^l\left({1\over x}{d\over dx}\right)^l\left({\e^{ix}\over
ix}\right)\eqno()$$
These functions satisfy recursion relations\footnote{**}{Note that for $D=4$,
\(gensol c) involves $h_{-1}$. In this case we define $h_{-1}\equiv ih_0$, so
that these relations work for $l=-1$ and $l=1$ respectively.}
$$\frac{d}{dx}h_l(x)-\frac lx h_l(x)=-h_{l+1}(x)\eqno(regurgitation a)$$
$$\frac{d}{dx}h_l(x)+\frac{l+1}xh_l(x)=h_{l-1}(x)\eqno(regurgitation b)$$
We then need to enforce the gauge condition $F_\mu=0$. In terms of the
polarization coefficients $A,\ B,\ C$ this condition becomes
$$i k^\ibar A_{\ibar\mubar}+i\sqrt{k^2+\alpha^2}A_{t\mubar}-\alpha B_\mubar
=0\eqno(cantthinkofagoodnameforthisone a)$$
and
$$\frac1{D-3}\alpha A_\mubar^\mubar+ik^\ibar B_\ibar+i\sqrt{k^2+\alpha^2}
B_t+\frac{D-2}{D-3}\alpha C=0\eqno(cantthinkofagoodnameforthisone b)$$
The space of solutions still has a residual gauge invariance under
transformations which preserve $F_\mu=0$. We could use this residual
invariance to solve \(cantthinkofagoodnameforthisone) and eliminate some
parameters. There are several ways to do this. One would be to set all the
$B$'s and $C$ to zero. Then we also need $A$ to be traceless, and to
satisfy $k^\ibar A_{\ibar\mubar}+\sqrt{k^2+\alpha^2}A_{t\mubar}=0$.
Alternatively we could set all timelike components zero: $A_{t\mubar}=
B_t=0$. Then we need to have \def\jbar{{\overline j}}
$B_\ibar=\frac i\alpha k^\jbar A_{\ibar\jbar}$ and $C=\frac1{D-2}\left(
-A_{\ibar\ibar}-\frac i\alpha(D-3)k^\ibar B_\ibar\right)$. This last solution
gives a Fock space of manifestly nonnegative norm.

These are the solutions for $F=0$. We can also ask if there are any solutions
with $F$ nonzero. In order that such solutions satisfy both the invariant
and fixed equations of motion, $F$ must satisfy
$$F_{(\mu,\nu)}-\frac12\eta_{\mu\nu}{F^\rho}_{,\rho}+\frac{D-2}{2y}
\delta_{(\mu}^yF_{\nu)}+\frac{D-2}{4y}\eta_{\mu\nu}F^y=0\eqno(Feqn)$$
The only solutions to this are
$$F_\mu=y^{1-D/2}\ \overline\delta_\mu^\mubar\left(c_{\mubar\nubar}x^\nubar
+d_\mubar\right)\eqno()$$
where $c$ is antisymmetric.
There are not any $\psi$ solutions of the free equations which are bounded for
$|x^\ibar|\rightarrow\infty$ and give such results for $F$.

\head{Retarded Green's Function}

The retarded Green's function for our theory must satisfy
$${D_{\mu\nu}}^{\rho\sigma}
\left[_{\rho\sigma}G^{\alpha\beta}_{{\rm ret}}\right](x,x')
=\delta_\mu^{(\alpha}\delta_\nu^{\beta)}\delta^D(x-x')\eqno()$$
It must also obey retarded boundary conditions. As we have discussed
previously, we need to extend AdS to the covering space CAdS. The time
coordinates $t,\ t'$ determine which is the earlier point only in the
case that the winding numbers $N,\ N'$ are equal. If $N<N'$ then $x$
corresponds to an earlier event than $x'$, and vice versa, regardless
of $t$ and $t'$. So the correct retarded boundary condition is
$$G(x,x')=0\hbox{ if }t<t'\hbox{ and }N=N'$$
and
$$G(x,x')=0\hbox{ if }N<N'$$
We will proceed to solve this explicitly shortly, but first we discuss in
general whether the solution to the {\it gauge-fixed} equations of
motion generated by
$$\psi_{\alpha\beta}(x)=\int d^Dx'
\left[_{\alpha\beta}G^{\mu\nu}_{{\rm ret}}\right](x,x')
T_{\mu\nu}(x')\eqno(Gsol)$$
must also solve the gauge-invariant equations of motion
$${\delta S_{Inv}\over\delta\psi^{\alpha\beta}}=T_{\alpha\beta}\eqno(inveqn)$$
when the source $T$ satisfies the appropriate conservation law. This law is
the requirement that
$$\int d^Dx\ T^{\alpha\beta}(x)\  \delta_e
\psi_{\alpha\beta}(x)=0\eqno(conslaw)$$
for any gauge parameters $e_\mu(x)$. If $T$ does not satisfy \(conslaw) then
no solutions exist to \(inveqn). Now consider the variation of the
gauge-fixed action
$$S_{Fixed}=S_{Inv}-\half \int d^Dx F_\mu(x) F^\mu(x)\eqno(sgf)$$
under a gauge transformation:
$$\int d^Dx\ (\delta_e\psi^{\alpha\beta}(x))\ {\delta S_{Fixed}\over\delta
\psi^{\alpha\beta}(x)}=0-\int d^Dx F_\mu(x)\ \delta_e F^\mu(x)\eqno(hwat)$$
Now evaluate this with $\psi$ set equal to the solution \(Gsol). Then
${\delta S_{Fixed}\over\delta\psi^{\alpha\beta}(x)}$ is $T_{\alpha\beta}(x)$
by the gauge-fixed equations of motion. Then the left side of \(hwat) must
be zero because $T$ is a conserved current. Then so too must the right side,
and integrating this by parts we see that we must have
$$\left(-\partial^2+{D(D-2)\over4y^2}\right)F_\mubar=0\eqno(dickweed a)$$
$$\left(-\partial^2+{(D-2)(D-4)\over4y^2}\right)F_y=0\eqno(dickweed b)$$
It is easy to see that \(Feqn) implies \(dickweed), but not the other way
around. If our solution also satisfies \(Feqn), then it will obey the
invariant as well as gauge fixed equations of motion.

We now move on to the explicit determination of $G$.
After some analysis which exactly parallels that of the de Sitter
case\refto{structure}, we determine the tensor structure of $G$:
$$\left[_{\rho\sigma}G^{\alpha\beta}_{{\rm ret}}\right](x,x')=
a(x,x')\left[_{\rho\sigma}T_A^{\alpha\beta}\right]+
b(x,x')\left[_{\rho\sigma}T_B^{\alpha\beta}\right]+
c(x,x')\left[_{\rho\sigma}T_C^{\alpha\beta}\right]\eqno()$$
where
$$\left[_{\rho\sigma}T_A^{\alpha\beta}\right]\equiv
2\overline\delta_{(\rho}^\alpha\overline\delta_{\sigma)}^\beta
-\frac2{D-3}\overline\eta^{\alpha\beta}\overline\eta_{\rho\sigma}\eqno(t a)$$
$$\left[_{\rho\sigma}T_B^{\alpha\beta}\right]\equiv 4\delta^y_{(\rho}
\overline\delta_{\sigma)}^{(\alpha}\delta_y^{\beta)}\eqno(t b)$$
$$\left[_{\rho\sigma}T_C^{\alpha\beta}\right]
\equiv\frac1{D-2}\left[\frac2{D-3}\overline\eta_{\rho\sigma}
\overline\eta^{\alpha\beta}-2\delta_\rho^y\delta_\sigma^y
\overline\eta^{\alpha\beta}-2\overline\eta_{\rho\sigma}
\delta^\alpha_y\delta^\beta_y+2(D-3)\delta_\rho^y\delta_\sigma^y
\delta^\alpha_y\delta^\beta_y\right]\eqno(t c)$$
and $a,b,c$ obey the equations
$$D_0\ a(x,x')=\delta^D(x-x')\eqno(whatever a)$$
$$D_{D-2}\ b(x,x')=\delta^D(x-x')\eqno(whatever b)$$
$$D_{2(D-3)}\ c(x,x')=\delta^D(x-x')\eqno(whatever c)$$
Because of the delta functions, we can write these equations
in the symmetric form
$$\left(yy'\right)^{{2-D\over2}}\left(\partial^2-{D-2\over y}{\partial
\over\partial y}+{n\over y^2}\right)G_n=\delta^D(x-x')\eqno()$$
To solve these equations explicitly, we expand the delta functions:
$$\delta^D(x-x')=\delta^{D-2}(x^\ibar-x'{}^\ibar)\ \delta(y-y')\ \delta(t-t')
\ \delta_{N,N'}\eqno(delexp a)$$
$$\delta^{D-2}(x^\ibar-x'{}^\ibar)=\left(\frac1{2\pi}\right)^{D-2}\int d^{D-2}k
\ \e^{i k^\ibar (x-x')^\ibar}\eqno(delexp b)$$
$$\delta(y-y')=\int_{-\infty}^\infty {d\alpha\over2\pi}\ \alpha^2\ yy'\ h_l
(\alpha y)\ h_l^*(\alpha y')\eqno(delexp c)$$
The expansion \(delexp c) works for any $l$, but we choose
$l=\sqrt{{1\over 4}(D-1)^2-n}$ to diagonalize $D_n$.
We then similarly expand $G_n$ as
$$G_n(x,x')=\int{d^{D-2}k\over (2\pi)^{D-2}}\int_{-\infty}^\infty
{d\alpha\over2\pi}\ \e^{ik^\ibar(x-x')^\ibar}\alpha^2\left(yy'\right)^{D/2}
h_l(\alpha y)h_l^*(\alpha y')\ g_n(t,t',\alpha,k^2)\eqno()$$
Then we find
$$\left(-{\partial^2\over\partial t^2}-k^2-\alpha^2\right)g_n=
\delta(t-t')\ \delta_{N-N'}\eqno()$$
We will first solve this for $N=N'$ and later generalize. This solution is
\def\ka{\sqrt{k^2+\alpha^2}}
$$g_n=-{\sin(\ka(t-t'))\over\ka}\theta(t-t')\eqno()$$
so
$$G_n=-\int{d^{D-2}k\ d\alpha\over(2\pi)^{D-1}}\ \e^{ik^\ibar(x-x')^\ibar}
\ \alpha^2(yy')^{D/2}\ {\sin(\ka(t-t'))\over\ka}\ h_l(\alpha y)h_l^*
(\alpha y')\ \theta(t-t')\eqno(Gnsol)$$
The evaluation of this integral depends strongly on the value of $D$. We
will now evaluate it for the physical case of $D=4$. To do so, we notice
that our theory is Lorentz invariant in the flat subspace $t,x^\ibar$.
If $(t-t')^2-(x-x')^\ibar(x-x')^\ibar<0$ then we can, without changing
the background metric, go to a frame where $t-t'=0$, in which case we
see that $G_n=0$. Otherwise, we can go to a frame where $x^\ibar-x'{}^\ibar
=0$. Then
$$G_n=-\int {d^2k\ d\alpha\over(2\pi)^3}\ \alpha^2(yy')^2\ {\sin(\ka(t-t'))
\over\ka}\ h_l(\alpha y)\ h_l^*(\alpha y')\ \theta(t-t')\eqno()$$
We write the $k$ integral in polar coordinates and substitute $u=\ka$ to
get
$$G_n=-{(yy')^2\over t-t'}\ \theta(t-t')\ \int{d\alpha\over(2\pi)^2}\ \alpha^2
\cos\alpha(t-t')\ h_l(\alpha y)h^*_l(\alpha y')\eqno()$$
Now we look at this for the $n$ values of interest. First $n=0$ gives
$l=1$. We get
$$G_0=-{yy'\over2(t-t')}\theta(t-t')\int {d\alpha\over(2\pi
)^2}\left(1+{i(y'-y)\over yy'\alpha}+{1\over\alpha^2}\right)\left(\e^{i\alpha
(t-t'+y-y')}+\e^{i\alpha(-t+t'+y-y')}\right)\eqno(sillyname)$$
The first term in \(sillyname) just gives delta functions. The others can
be evaluated by contour integration, and we get
$$G_0=\frac1{4\pi}\theta(t-t')\left[\theta(t-t'-|y-y'|)-{yy'\over t-t'}
\left(\delta(t-t'+y-y')+\delta(t-t'-y+y')\right)\right]\eqno()$$
Now we need to generalize this result to $x^\ibar-x'{}^\ibar\ne 0$ and
$N\ne N'$. The straightforward invariant and causal generalization is
$$G_0(x,x')=\frac1{4\pi}\Biggl\lbrace\left[\theta(t-t')\delta_{N,N'}+
\theta(t'-t)\delta_{N,N'+1/2}\right]\left[\theta(1-z(x,x'))-\half
\delta(1-z(x,x'))\right]$$
$$+\theta(t-t')\delta_{N,N'+1/2}+
\theta(N-N'-\frac34)\Biggr\rbrace\eqno(gzero)$$
where we recall that $1-z(x,x')$ is given by \(confdist). \(gzero) is
constructed following the discussion at the end of the second section. The
delta function term is a massless signal so it only propagates to the next
half level. The theta function term is a timelike signal so its
influence is felt on the next half level $N=N'+\half$ at all times such
that $t>t'-\sqrt{(x-x')^i(x-x')^i}$, and thereafter at all locations for
$N\ge N'+1$ (hence the final term with no theta function). If we had used
some sort of reflective boundary conditions then the delta function would
need to be continued to all $N>N'$ (possibly with an alternating sign,
depending on which type of boundary conditions were used).

So much for the case of $n=0$. The other two cases are the same $(n=2)$ in
$D=4$. For $N=N'$ and $x^\ibar=x'{}^\ibar$ they evaluate to
$$G_2=-{yy'\over4\pi(t-t')}\theta(t-t')\left[\delta(t-t'+y-y')+\delta(-t+t'
+y-y')\right]\eqno()$$
which generalizes to
$$G_2(x,x')=-\frac1{8\pi}\left[\theta(t-t')\delta_{N,N'}+\theta(t'-t)
\delta_{N,N'+1/2}\right]\delta(1-z(x,x'))\eqno()$$
Putting it all together with the tensor factors, the final answer for our
Greens function is
$$\left[_{\rho\sigma}G^{\alpha\beta}\right](x,x')=\frac1{4\pi}\Biggl\lbrace
\left[\theta(t-t')\delta_{N,N'}+\theta(t'-t)\delta_{N,N'+1/2}\right]
\Biggl[\theta(1-z(x,x'))\left(2\overline\delta_\rho^{(\alpha}
\overline\delta_\sigma^{\beta)}-2\overline\eta_{\rho\sigma}
\overline\eta^{\alpha\beta}\right)$$
$$-\half\delta(1-z(x,x'))
\left(2\delta_\rho^{(\alpha}\delta_{\sigma}^{\beta)}
-\eta_{\rho\sigma}\eta^{\alpha\beta}\right)
\Biggr]\eqno(bigG)$$
$$+\left[\theta(t-t')\delta_{N,N'+1/2}+\theta(N-N'-\frac34)\right]
\left(2\overline\delta_\rho^{(\alpha}\overline\delta_{\sigma}^{\beta)}
-2\overline\eta_{\rho\sigma}\overline\eta^{\alpha\beta}\right)\Biggr\rbrace$$
In the next section
we show that this Greens function gives the correct anti de
Sitter-Schwarzchild solution as its response to a single point mass.
\endpage

\head{Response to a Point Mass}

Consider a point mass which is at rest at $r=0$ in the static coordinates
\(smet). These coordinates are related to our conformal coordinates as
follows:
\def\cowcrap#1{\left(r\cos\theta+\sqrt{1+r^2}\cos\tau\right)^{#1}}
\def\crap{\cowcrap{-1}}
$$y=\crap\eqno(crap a)$$
$$t=\sqrt{1+r^2}\sin\tau\crap\eqno(crap b)$$
$$x^1=r\sin\theta\cos\phi\crap\eqno(crap c)$$
$$x^2=r\sin\theta\sin\phi\crap\eqno(crap d)$$
We see that a point mass at $r=0$ has $x^1=x^2=0$, and $y^2=t^2+1$. So the
path is $y=\pm\sqrt{t^2+1}$, the sign depending on whether we are on an
$N=$ integer level $(y>0)$ or $N=$ half-integer level $(y<0)$. The action
for a point particle of mass $M$
which follows a spacetime path $q^\mu(\tau)$ is
$$S=-M\int d\tau \sqrt{-g_{\mu\nu}\dot q^\mu \dot q^\nu}\eqno()$$
The linearized energy-momentum tensor for the particle is then
$$T^{\alpha\beta}(x)=-{\delta S\over\delta\psi_{\alpha\beta}(x)}
\Bigr|_{\psi=0}\eqno()$$
Calculating this for our particle, we find the nonzero components of $T$ are
$$T^{tt}(x)=\mp\half\kappa M\ \delta^2(x^\ibar)\ \delta(y\mp\sqrt{t^2+1})
\eqno(teafortwo a)$$
$$T^{ty}(x)=-\half\kappa M{t\over\sqrt{t^2+1}}\ \delta^2(x^\ibar)
\ \delta(y\mp\sqrt{t^2+1})\eqno(teafortwo b)$$
$$T^{yy}(x)=\mp\half\kappa M{t^2\over t^2+1}\ \delta^2(x^\ibar)
\ \delta(y\mp\sqrt{t^2+1})\eqno(teafortwo c)$$
This $T$ must satisfy the conservation law which we get from \(conslaw):
$${T^{\mu\nu}}_{,\nu}+\frac1y\delta^\mu_yT^\nu_\nu=0\eqno(conslaw2)$$
It is easily verified that \(conslaw2) is in fact obeyed by \(teafortwo).

We now wish to explicitly compute the response to the source $T(x)$:
$$\kappa\psi_{\rho\sigma}(x)=\kappa \int d^4x'\left[_{\rho\sigma}
G^{\alpha\beta}\right](x,x')T_{\alpha\beta}(x')\eqno(response)$$
We can tell by the forms of \(bigG) and \(teafortwo) that this response
will have the form
$$\kappa\psi_{\alpha\beta}(x)=\eta_{\alpha\beta}A(x)+\delta_\alpha^t
\delta_\beta^tB(x)+\delta_\alpha^y\delta_\beta^yC(x)+\left(\delta_\alpha^t
\delta_\beta^y+\delta_\alpha^y\delta_\beta^t\right)D(x)\eqno(psiexp)$$
Now we need to evaluate the functions $A$ through $D$. For definiteness
we will take both $y$ and $t$ to be positive. We need to look at each term
in \(bigG) seperately: First,
$$\left[_{\rho\sigma}G^{\alpha\beta}_1\right](x,x')=\frac1{4\pi}
\left[\theta(t-t')\delta_{N,N'}+\theta(t'-t)\delta_{N,N'+1/2}\right]
\theta(1-z(x,x'))\left(2\overline\delta_\rho^{(\alpha}
\overline\delta_\sigma^{\beta)}-2\overline\eta_{\rho\sigma}
\overline\eta^{\alpha\beta}\right)\eqno(g1)$$
$G_1$ annihilates all components of $T$ except $T^{tt}$. Thus its
contribution will be of the form $\overline\eta_{\rho\sigma}+\delta_\rho^t
\delta_\sigma^t$. So $A_1=B_1=-C_1$, and $D_1=0$. Then we
calculate
$$A_1=-\frac1{4\pi}\kappa^2 M\int dt'\left[
\theta(t-t')\theta(2y\sqrt{t'{}^2+1}-2tt'-l)
-\theta(-t+t')\theta(-2y\sqrt{t'{}^2+1}+2tt'+l)\right]\eqno(steaksauce)$$
where
$$l\equiv -t^2+x^\ibar x^\ibar+y^2+1\eqno()$$
\(steaksauce) evaluates to
$$A_1=-4MG\left({lt-ys\over2(y^2-t^2)}+t\right)\eqno()$$
with
$$s\equiv\sqrt{l^2+4(t^2-y^2)}\eqno(sdef)$$
Note that \(sdef) gives a real $s$ for any $t$ and $y$. Now for the second
part of $G$, which is:
$$\left[_{\rho\sigma}G^{\alpha\beta}_2\right](x,x')=-\frac1{8\pi}
\left[\theta(t-t')\delta_{N,N'}+\theta(t'-t)\delta_{N,N'+1/2}\right]
\delta(1-z(x,x'))
\left(2\delta_\rho^{(\alpha}\delta_{\sigma}^{\beta)}
-\eta_{\rho\sigma}\eta^{\alpha\beta}\right)
\eqno(g2)$$
When we contract this with \(teafortwo), we get
\goodbreak
$$\frac{M\kappa^2}{2\pi}y\int dt' \sum_{\pm}\theta(\pm(t-t'))\ \delta(\pm2y
\sqrt{t'{}^2+1}-2tt'-l)$$
\nobreak
$$\times\left[\pm\left(\delta_\rho^t\delta_\sigma^t+\half\eta_{\rho\sigma}
\right)\sqrt{t'{}^2+1}\pm\left(\delta_\rho^y\delta_\sigma^y-\half\eta_{\rho
\sigma}\right){t'{}^2\over\sqrt{t'{}^2+1}}-\left(\delta_\rho^t\delta_\sigma^y
+\delta_\rho^y\delta_\sigma^t\right)\right]\eqno()$$
The delta function for the minus sign is not zero for any $t'$ values for
which $\theta(-t+t')\ne 0$. For the plus sign, there is one such value, which
is
$$t'_0={lt-ys\over 2(y^2-t^2)}\eqno(alexhaley)$$
Then we use the identity
$$\delta(f(t'))=\left|{df\over dt'}\right|^{-1}_{t'=t'_0}\delta(t')
\eqno()$$
to find the results
$$A_2=4MG{y\over s}\eqno(partdeux a)$$
$$B_2=4MG{y(ly-ts)^2\over2s(y^2-t^2)^2}\eqno(partdeux b)$$
$$C_2=-2A_2+B_2\eqno(partdeux c)$$
$$D_2=-4MG{y(lt-ys)(ly-ts)\over2s(y^2-t^2)^2}\eqno(partdeux d)$$
Finally, there is the third term in the Greens function
$$\left[_{\rho\sigma}G_3^{\alpha\beta}\right](x,x')=\frac1{4\pi}
\left[\theta(t-t')\delta_{N,N'+1/2}+\theta(N-N'-\frac34)\right]
\left(2\overline\delta_\rho^{(\alpha}\overline\delta_{\sigma}^{\beta)}
-2\overline\eta_{\rho\sigma}\overline\eta^{\alpha\beta}\right)\eqno(g3)$$
This gives a contribution of the form
\def\T{{\cal T}}
$$-4mG\left(\overline\eta_{\rho\sigma}+\delta_\rho^t\delta_\sigma^t
\right)\left(\T-t\right)\eqno()$$
The constant $\T$ is formally divergent, but this divergence does not affect
any physical observables. This is because it can be removed by a simple
rescaling of the flat spatial coordinates: $x^\ibar\rightarrow (1+2\kappa^2
\T)x^\ibar$. This divergence is a result of the unphysical assumption that
our space was anti de Sitter since infinitely long ago. If we had used
reflective boundary conditions, there would have been additional divergent
terms whose forms would be similar
to \(partdeux). These do not seem to be removable by coordinate
transformations.

For the form \(psiexp), we calculate
$$\kappa F_\mu=\frac1y\Biggl[(-A+\half B-\half C)_{,\mu}+\delta_\mu^y
\left(-\frac2yA+C_{,y}-\frac2yC-D_{,t}\right)+\delta_\mu^t\left(-B_{,t}+D_{,y}
-\frac2yD\right)\Biggr]\eqno()$$
which evaluates to
$$\kappa F_\mu={4mG\over y}\delta_\mu^t\eqno()$$
We can see that this $F$, while not zero, satisfies the equation \(Feqn). Thus
our solution will obey the gauge invariant equations of motion. One can check
this directly by computing the first-order
Riemann tensor for the solution \(psiexp):
\def\mbar{{\overline m}}\def\kbar{{\overline k}}\def\lbar{{\overline l}}
$${R^t}_{yty}=-\frac1{y^2}+\frac1{2y^2}\left(-y(B+C)_{,y}+y^2(-A+B)_{,yy}
+2yD_{,t}-2y^2D_{,ty}+y^2(A+C)_{,tt}\right)\eqno(reem a)$$
$${R^t}_{y\ibar\jbar}=0\eqno(reem b)$$
$${R^t}_{yt\ibar}=\frac1{2y}\left(-A+C+y(A-B)_{,y}+yD_{,t}\right)_{,\ibar}
\eqno(reem c)$$
$${R^t}_{yy\ibar}=\frac1{2y}\left(-D-yD_{,y}+y(A+C)_{,t}\right)_{,\ibar}
\eqno(reem d)$$
$${R^t}_{\ibar t\jbar}=-\frac1{y^2}\overline\delta_{\ibar\jbar}+\half
(-A+B)_{,\ibar\jbar}+\frac1{2y^2}\left(2C+2yA_{,y}-yB_{,y}+2yD_{,t}+y^2
A_{,tt}\right)\overline\delta_{\ibar\jbar}\eqno(reem e)$$
$${R^t}_{\ibar y\jbar}=\half D_{,\ibar\jbar}+\frac1{2y}\left(A_{,t}
+C_{,t}+yA_{,ty}\right)\overline\delta_{\ibar\jbar}\eqno(reem f)$$
$${R^t}_{\ibar\jbar\kbar}=\half\overline\epsilon_{\jbar\kbar}
\overline\epsilon_{\ibar\lbar}
\left(-\frac1yD+A_{,t}\right)_{,\lbar}\eqno(reem g)$$
$${R^y}_{\ibar y\jbar}=-\frac1{y^2}\overline\delta_{\ibar\jbar}
-\half(A+C)_{,\ibar\jbar}+\frac1{2y^2}\left(2C-yC_{,y}-y^2A_{,yy}\right)
\overline\delta_{\ibar\jbar}\eqno(reem h)$$
$${R^y}_{\ibar\jbar\kbar}=\half\overline\epsilon_{\jbar\kbar}
\overline\epsilon_{\ibar\lbar}
\left(\frac1yA+\frac1yC+A_{,y}\right)_{,\lbar}\eqno(reem i)$$
$${R^\ibar}_{\jbar\kbar\lbar}=\frac1{y^2}\left(-1+C-\frac{y^2}2A_{,\mbar\mbar}
+yA_{,y}\right)\overline\epsilon_{\ibar\jbar}\overline\epsilon_{\kbar\lbar}
\eqno(reem j)$$
where, as one might expect, $\overline\epsilon_{12}=-\overline\epsilon_{21}
=1,\ \overline\epsilon_{11}=\overline\epsilon_{22}=0.$
{}From \(reem), we obtain the Ricci tensor
$$R_{tt}=\frac3{y^2}-\frac3{y^2}(B+C)+\half\partial^2(A-B)+\frac1{2y}(-4A
+C+3B)_{,y}-\frac3yD_{,t}+D_{,ty}-\half(2A+B+C)_{,tt}\eqno(ricky a)$$
$$R_{yt}=-\frac3{y^2}D-\half\overline\delta^{\ibar\jbar}D_{,\ibar\jbar}
-\frac1y(A+C)_{,t}-A_{,ty}\eqno(ricky b)$$
$$R_{yy}=-\frac3{y^2}-\half\partial^2(A+C)-\frac1{2y}(3C+B)_{,y}+\half
(B+C)_{,yy}+\frac1yD_{,t}-D_{,ty}\eqno(ricky c)$$
$$R_{t\ibar}=\left(-\frac1y D+\half D_{,y}-A_{,t}-\half C_{,t}
\right)_{,\ibar}\eqno(ricky d)$$
$$R_{y\ibar}=\left(-\frac1yA-\frac1yC-A_{,y}+\half B_{,y}-\half D_{,t}
\right)_{,\ibar}\eqno(ricky e)$$
$$R_{\ibar\jbar}=-\frac3{y^2}\overline\delta_{\ibar\jbar}
+\left(-A-\half C+\half B
\right)_{,\ibar\jbar}+\left(-\half\partial^2 A+\frac3{y^2}C+\frac1{2y}(4A
-B-C)_{,y}+\frac1yD_{,t}\right)\overline\delta_{\ibar\jbar}\eqno(ricky f)$$
Substituting the explicit values for the coefficient functions, we can show
$$R_{\mu\nu}=-\frac3{y^2}\left(\eta_{\mu\nu}+\kappa\psi_{\mu\nu}\right)
\eqno()$$
to linearized order.

This shows that the point mass response solves Einstein's equations. Now
we need to determine whether the solution generated is equivalent to the anti
de Sitter-Schwarzchild solution. In static spherical coordinates $\tau,\ r
,\ \theta,\ \phi$, the anti de Sitter-Schwarzchild metric is
$$g_{\tau\tau}=-1-r^2+{2MG\over r}\eqno(ashole a)$$
$$g_{rr}=\left(1+r^2-{2MG\over r}\right)^{-1}={1\over 1+r^2}
+{2MG\over r(1+r^2)^2}+O(G^2)\eqno(ashole b)$$
$$g_{\theta\theta}=r^2\eqno(ashole c)$$
$$g_{\phi\phi}=r^2\sin^2\theta\eqno(ashole d)$$
with all off-diagonal components of $g$ zero.
Does a coordinate transformation exists from $\tau,\ r,\ \theta,\ \phi$ to
$t,\ x^1,\ x^2,\ y$ such that \(ashole) is transformed into our point mass
solution? If so, then it is given by \(crap) to lowest order, but there will
will be order $MG$ corrections to \(crap) which we don't know. This
ambiguity in the order $MG$ coordinate transformations is related to the
fact that we had a choice of gauge in solving for our Greens function.
Making order $MG$ coordinate transformations on the point mass solution will
give the same solution in a different gauge.
It would be quite difficult to find the exact transformation which
would give the solution in our gauge.

It is easier to show indirectly that our solution is equivalent to \(ashole).
We can do this by comparing the Weyl tensors for both solutions. This is
given in general by \(vile); for a metric such as ours which obeys Einstein's
equations with $\Lambda=-3$, this is
$$C^\lambda{}_{\mu\nu\kappa}=R^\lambda{}_{\mu\nu\kappa}+\delta^\lambda_\nu
g_{\mu\kappa}-\delta^\lambda_\kappa g_{\mu\nu}\eqno(morevile)$$
Evaluating this for our solution, we find
$$C^t{}_{y12}=0\eqno(sovile a)$$
$$C^t{}_{1t2}=96MG\ {x_1x_2y(t^2+y^2)\over s^5}\eqno(sovile b)$$
$$C^t{}_{1y2}=-192MG\ {x_1x_2ty^2\over s^5}\eqno(sovile c)$$
$$C^t{}_{112}=-48MG\ {x_2ty(1+t^2+x_1^2+x_2^2-y^2)\over s^5}\eqno(sovile d)$$
$$C^y{}_{112}=-48MG\ {x_2y^2(1+t^2+x_1^2+x_2^2-y^2)\over s^5}\eqno(sovile e)$$
$$C^y{}_{1y1}=-8MGy{s^2+12t^2x_1^2-12y^2x_2^2\over s^5}\eqno(sovile f)$$
These are the independent components of $C$; the other components can be
obtained from these by switching $x_1\leftrightarrow x_2$, using the standard
permutation symmetries of $C$, or using the
fact that contracting any two indices of $C$ with the
metric gives zero. Compare this with the first-order Weyl tensor obtained from
the anti de Sitter-Schwarzchild solution:
$$C^{r\tau r\tau}=-{2MG\over r^3}\eqno(revile a)$$
$$C^{\theta\tau\theta\tau}={MG\over(r^2+1)r^5}\eqno(revile b)$$
$$C^{\theta r\theta r}=-{MG(r^2+1)\over r^5}\eqno(revile c)$$
$$C^{\phi r\phi r}=-{MG(r^2+1)\over r^5\sin^2\theta}\eqno(revile d)$$
$$C^{\phi\tau\phi\tau}={MG\over r^5(r^2+1)\sin^2\theta}\eqno(revile e)$$
$$C^{\phi\theta\phi\theta}={2MG\over r^7\sin^2\theta}\eqno(revile f)$$
We have written $C$ with all indicies up, to facilitate transforming it to
conformal coordinates. The transformed $C$ is $C'$, where
\def\px#1#2{{\partial x'{}^#1\over\partial x^#2}}
$$C'^{\alpha\beta\gamma\delta}=\px\alpha\mu\px\beta\nu\px\gamma\rho
\px\delta\sigma C^{\mu\nu\rho\sigma}\eqno()$$
where $x^\mu=(\tau,\ r,\ \theta,\ \phi)$, and $x'{}^\alpha=
(t,\ x_1,\ x_2,\ y)$. Because $C$ is zero for the background anti de Sitter
metric, we need only use the zero-order transformations \(crap). The results
of this are
$$C^{ty12}=0\eqno(vilecabinet a)$$
$$C^{t1t2}=-{3MG\cos\phi\sin\phi\sin^2\theta(1+\sin^2\tau+r^2\sin^2\tau)
\over r^3\cowcrap6}\eqno(vilecabinet b)$$
$$C^{t1y2}=-{6MG\sin\phi\cos\phi\sin^2\theta\sin\tau\sqrt{1+r^2}\over
r^3\cowcrap6}\eqno(vilecabinet c)$$
$$C^{t112}=-{3MG\sin\phi\sin\tau\sin\theta(r\sqrt{1+r^2}+\cos\tau\cos\theta
(r^2+1))\over r^3\cowcrap6}\eqno(vilecabinet d)$$
$$C^{y112}=-{3MG\sin\phi\sin\theta(r+\cos\tau\cos\theta
\sqrt{r^2+1})\over r^3\cowcrap6}\eqno(vilecabinet e)$$
$$C^{y1y1}=-{MG\over r^3}\left[{1\over\cowcrap3}+3\sin^2\theta{\sin^2\phi
+(r^2+1)\sin^2\tau\cos^2\phi\over\cowcrap6}\right]
\eqno(vilecabinet f)$$
We then invert \(crap) to find
$$r={s\over 2|y|}\eqno(parc a)$$
$$\tau=\tan^{-1}{2t\over l}\eqno(parc b)$$
$$\theta=\tan^{-1}{2\sqrt{x_1^2+x_2^2}\over 2-l}\eqno(parc c)$$
$$\phi=\tan^{-1}{x_2\over x_1}\eqno(parc d)$$
Substituting these expressions into \(vilecabinet), and using the background
metric $y^{-2}\eta_{\alpha\beta}$ to lower the last three indices, we find
that \(vilecabinet) is equal to \(sovile). Thus, our solution is just the
anti de Sitter-Schwarzchild solution in different coordinates.

\endpage

\head{Conclusions}

We have determined the retarded Greens function
for quantum gravity in the idealized case of a background which is anti de
Sitter at all locations and all times. To do so, we needed
to establish a boundary condition at spatial infinity (which corresponds
to $y=0$ in our conformal coordinate system). We mandated the boundary
condition that any signal which reaches spatial infinity should be lost,
not reflected back. We showed that this boundary condition gives the
correct classical solution for the case of a point mass at the origin,
up to of course a change of coordinates.

It is easy to go from the Greens function \(bigG) to the Feynman propagator
for gravity in anti de Sitter space for the case $N=N'$, up to real analytic
terms which are annihilated by the kinetic operator. This will be
$$i\left[_{\rho\sigma}\Delta^{\alpha\beta}\right](x,x')=\frac1{16\pi^2}
\Biggl\{{1\over 1-z(x,x')+i\epsilon}-\ln\left[(x-x')^2+i\epsilon\right]
\Biggr\}\eqno()$$
As in the de Sitter case\refto{structure,me}, the propagator cannot be made
invariant under the anti de Sitter symmetry.

For $N\ne N'$ it is not clear
how to obtain the propagator, because of the other terms in
the Greens function. Loss of quantum coherence may be a problem
here, since information which reaches spatial infinity ($y=0$ in
conformal coordinates) will be lost. We still advocate the boundary
conditions we have used here rather than reflective ones, because as we have
shown, our boundary conditions give the correct classical solution (at least
for one specific case). This loss of coherence will not be
present when we go to physically realistic situations. If it occurs in the
case of infinite eternal anti de Sitter space, then it just serves to
underscore the unphysiciality of this mathematical model.

To proceed further, we would need to understand what modifications of this
work would be necessary so that it might apply to a physically realistic
situation. A negative cosmological constant might have existed in the
physical universe as the result of a field theoretic phase transition
associated with a symmetry breaking. Such a phase transition alters the
effective vacuum energy by redefining the vacuum. There are two differences
between this realistic situation and our mathematical model. First, by
causality the phase transition cannot occur throughout the whole universe at
one time. Therefore a region of anti de Sitter space would be surrounded by
normal space. Secondly, the universe would not have been anti de Sitter
for all time, as our model space in this paper was.
These features will likely make dealing with a realistic anti de Sitter
phase more difficult than the simple mathematical model which we have solved
in this paper.

\head{Acknowledgements}

The author thanks R. Woodard for suggesting this project, and for discussions
related to this work. This author also gratefully acknowledges support by
the United States Department of Energy under grant DE-FG05-84ER40141.
Some of the caucluations in the last section of this paper were done using
the symbolic manipulation program {\it Mathematica}.

\references

\refis{me}G. Kleppe, \pl B317, 1993, 305.

\refis{structure}N. C. Tsamis and R. P. Woodard, ``The Structure of
Perturbative
Quantum Gravity on a De Sitter Background'', Florida preprint UFIFT-92-14, to
appear in {\sl Comm. Math. Phys.}

\refis{mode}N. C. Tsamis and R. P. Woodard, \pl B292, 1992, 269.

\refis{strong}N. C. Tsamis and R. P. Woodard, \pl B301, 1993, 351; ``Strong
Infrared Effects in Quantum Gravity'', UFIFT-HEP-92-24/CRETE-92-17.

\refis{weinberg}S. Weinberg, \rmp 61, 1989, 1.

\refis{rentacar}S. J. Avis, C. J. Isham, and D. Storey, \prd 18, 1978, 3565.

\refis{whosonfirst}L. Abbott, {\sl Sci. Am.} {\bf 258} (1988) 106.

\refis{ramond}J. Minahan, P. Ramond, and R. Warner, \prd 41, 1990, 715.

\refis{getphysical}N. C. Tsamis and R. P. Woodard,
UFIFT-93-17/CRETE-93-11.

\endreferences

\end